\documentstyle[12pt]{article}

\input tcilatex

\begin{document}

\title{Spectral and thermodynamical properties of systems with noncanonical
commutation rules: semiclassical approach}
\author{J. C. Flores$^{\dagger }$ and S. Montecinos$^{\ddagger }$}
\date{$^{\dagger }$Universidad de Tarapac\'a, Departamento de F\'\i sica, Casilla
7-D, Arica, Chile.\\
$^{\ddagger }$Universidad de la Frontera, Departamento de F\'\i sica,
Casilla 54-D, Temuco, Chile.}
\maketitle

\begin{abstract}
We study different quantum one dimensional systems with noncanonical
commutation rule $[x,p]=i\hbar (1+sH),$ where $H$ is the one particle
Hamiltonian and $s$ is  a parameter. This is carried-out using semiclassical
arguments and the surmise $\hbar \rightarrow \hbar (1+sE),$ where $E$ is the
energy. We compute the spectrum of the potential box, the harmonic
oscillator, and a more general power-law potential $\left| x\right| ^{\nu }$%
. With the above surmise, and changing the size of the elementary cell in
the phase space, we obtain an expression for the partition function of these
systems. \ We calculate the first order correction in $s$ for the internal
energy and heat capacity. We apply our technique to the ideal gas, the
phonon gas, and to $N$ non-interacting particles with external potential
like $\left| x\right| ^{\nu }$.
\end{abstract}

\bigskip

\baselineskip=16pt

PACS and Keywords

\vskip 0.1cm 05.30.-d quantum statistical mechanics

03.65.-w quantum mechanics

03.65.Sq semiclassical theories

05.70.-a thermodynamics

\begin{section}*{I  Introduction}
\end{section}

In one dimension, some authors [1-9] have showed that a modification of the
usual commutation rule for the momentum and position operators: 
\begin{equation}
\lbrack x,p]=i\hbar \{1+sH(x,p)\},
\end{equation}
with $s$ a parameter and $H$ the Hamiltonian of the system, produces new
phenomena which appear, for instance, in high energy physics and mesoscopic
systems. In reference [1] it was shown that for some commutation relation
like (1), space discreteness is compatible with Lorentz transformation. This
fact was explicitly related to atomic phenomena. In [2,3] the mass-spectrum
for elementary particles was obtained from (1), with $H$ the Hamiltonian of
the harmonic oscillator and, applied for energy of the order of GeV-Tev (10$%
^{9}-10^{12}$ eV). In reference [4] it was shown that, for the free particle
Hamiltonian, (1) produces space quantization. This result can be related
with quark confinement phenomena. Moreover, in [5,6] mathematical aspects of
(1) were studied. In [7-10] it was found that charge discreteness in
mesoscopic circuits can be mathematically formulated with commutation
relations similar to (1) between charge and current. This theory becomes
related to the description of phenomena like persistent current in a ring of
inductance $L$, Coulomb blockage phenomena in a pure capacitor-design [7],
or current magnification [10]. In these cases the parameter $s$ becomes
related to the elementary charge $q_{e}$ by the relationship $%
s=Lq_{e}^{2}/\hbar ^{2}$. The analogy between [4] and [7] becomes from
charge and space quantization. We remark that in reference [2] the general
case $[x,p]=i\hbar f(H)$ \ was considered. The particular case $f=e^{sH}$
deserves some attention in our paper. In fact, thermodynamical properties
are easily found in this case.

\[
{} 
\]

In this article we propose a method, supported with semiclassical arguments,
which permits to calculate the spectrum of systems with canonical
commutation relations like (1). In fact, (1) suggests a formal energy \
dependence of the Planck constant $\hbar $ given by

\begin{equation}
\hbar \rightarrow \hbar \{ 1+sE\} .
\end{equation}

With this surmise we reproduce the spectrum of the systems treated in [2-4]
such as the harmonic oscillator, the free particle and the particle in a
quantum box. Moreover, we characterize the spectrum for the power-law
potential given by $U(x)=k|x|^{\nu }$.

\[
\]

On the other hand, we compute the thermodynamical properties of systems with
noncanonical commutation rule (1). The surmise (2) introduces a new size for
the elementary cell in the corresponding phase-space, and then for the
number of microstates. We calculate the internal energy and the heat
capacity for the ideal gas, for particles in a general potential $U\sim \mid
x\mid ^{\nu }$ and, finally, the phonon gas. We remark that thermodynamical
properties of a mesoscopic circuit, with charge discreteness [7-10], could
be studied by analogy with the mechanical case.

\[
{} 
\]

The paper is structured as follows: in section II, we compute the spectrum
of the harmonic oscillator and the potential box, in accordance with the
results found in the literature. Moreover, we consider systems with
potentials of the form $U\sim |x|^{\nu },~\nu >0$ and noncanonical
commutation rule (1). In section III, we study the thermodynamical
properties of these systems. We calculate the density of states and, at
first order in $s$, the internal energy and heat capacity. Finally, in
section IV, we give some conclusions and discussion.

\[
{} 
\]

\begin{section}*{II  Spectral properties: semiclassical approach}
\end{section}

The old quantum theory (Bohr-Wilson-Sommerfeld) provides a general method
for calculating spectra for some regular systems using simple rules. It
works for different systems like the harmonic oscillator, the hydrogen atom,
and others. In one dimension, the correspondence between classical and
quantum theory is established by the non linear differential equation
[11,12]:

\begin{equation}
\frac{dE}{dn}=\hbar \omega _{cl}(E),
\end{equation}
where $\omega _{cl}(E)$ denotes the classical frequency of a given system
with energy $E$. In (3), $\hbar $ is the Planck's constant, and $n$ will
correspond to the quantum number. The quantization of energy results from
the fact that $n$ can take only integer values in the solution of (3). The
initial condition $E(n=0)=E_{o}$ corresponds to the fundamental level of
energy. The deduction of (3) comes from the classical relation between the
action variable $I$ and the period of the classical orbit $T_{cl}$ where $%
\frac{dI}{dE}=T_{cl}$ and the quantization rule $I=n2\pi \hbar $.

\[
{} 
\]

Our hypothesis is that, for systems with noncanonical commutation rules (1),
the equation (3) must be modified by considering the surmise (2), namely

\begin{equation}
\frac{dE}{dn}=\hbar (1+sE)\omega _{cl}(E).
\end{equation}

Hereafter, we assume that $(1+sE)\geq 0$. Note that (4) introduces a new
fixed-point $E_{f}=-1/s$ which modifies the spectral structure of the
system. Since $\omega _{cl}>0$, the stability of this new fixed-point, with
respect to variations of $n$, \ depends only on the sign of the parameter $s$%
.

\[
{} 
\]

As a first application of the semiclassical equation (4), we consider a
particle of mass $m$ in a quantum box potential defined by

\begin{equation}
U(x)=\{ 
\begin{array}{cc}
0 & \mbox{if }0<x<a \\ 
\infty & \mbox{otherwise}~.
\end{array}
\end{equation}
The classical frequency is given by $\omega _{cl}=\frac \pi a\sqrt{\frac{2E}m%
}$ and the equation (4) becomes

\begin{equation}
\frac{dE}{dn}=\hbar \{1+sE\}\frac{\pi }{a}\sqrt{\frac{2E}{m}}.
\end{equation}
For $s>0$ the solution of (6) is given by 
\begin{equation}
E_{n}=\frac{1}{s}\tan ^{2}(\gamma n\sqrt{s}),
\end{equation}
where $\gamma =\frac{\pi \hbar }{\sqrt{2m}a}$, in accordance with the result
of reference [4]. The \ above result shows the power of our method because
the quantum calculation [4] \ is hard. Moreover (7) corresponds to one case
of finite number of states (when $\ \ \gamma \sqrt{s}=\pi /m$, $m$ integer).
Namely, our method also applies to these cases.

\[
{} 
\]

As a second application we consider the harmonic oscillator $U(x)=\frac
12\omega _o^2x^2$, where $\omega _o$ is the classical frequency. The
equation (4) for the spectrum becomes

\begin{equation}
\frac{dE}{dn}=\hbar \{1+sE\}\omega _{o}.
\end{equation}
The solution of this equation is 
\begin{equation}
E_{n}=(E_{o}+\frac{1}{s})e^{s\hbar \omega _{o}n}-\frac{1}{s},
\end{equation}
where the fundamental level $E_{o}=\hbar \omega _{o}/2$. This result is in
agreement, to first order in $s\hbar \omega _{o}$, with reference [3]. Note
that the spectrum is unbounded when $s>0$. Besides, if $-1/E_{o}<s<0$, then
in the limit $n\rightarrow \infty $ we have $E_{n}\rightarrow -1/s$ that
corresponds to the stable fixed point of equation (8).

\[
{} 
\]

Now, we consider the general case of a particle in the potential energy 
\begin{equation}
U(x)=k\mid x\mid ^{\nu },
\end{equation}
where $\nu $ and $k$ are arbitrary positive constants. In this case the
system admits always classically bounded trajectories and the classical
frequency is [13] :

\begin{equation}
\omega _{cl}(E)=\alpha (k,\nu )E^{\frac{1}{2}-\frac{1}{\nu }};\text{ where \
\ }\alpha (k,\nu )=\frac{\sqrt{2\pi }\nu k^{1/\nu }\Gamma (1/2+1/\nu )}{2%
\sqrt{m}\Gamma (1/\nu )},
\end{equation}
with $\Gamma $ the Gamma function. The differential equation (4) becomes

\begin{equation}
\frac{dE}{dn}=\hbar \alpha (k,\nu )\{1+sE\}E^{\frac{1}{2}-\frac{1}{\nu }},
\end{equation}
which could be analytically solved in some cases. Notice that (12) defines
in a direct way the density of states $dn/dE$ of these systems. This
expression will be re-obtained in the next section by an adequate definition
of the elementary cell volume in phase space and its thermodynamical
properties.

\[
\]

\begin{section}*{III  Thermodynamical properties}
\end{section}

In this section we compute thermodynamical properties of systems with
noncanonical commutation rules like (1). The change in $\hbar $ (2)
introduces a modification of the size of the elementary cell in the phase
space and then in the number of microstates. From the number of microstates,
we obtain the density of states and the partition function. As usual, the
knowledge of the partition function allows us to compute the thermodynamical
properties of the system. In particularly, we study the ideal gas, the
phonon gas, and then we generalize these results to the case with potential
(10).

\[
{} 
\]

In statistical mechanics [14], the number of microstates $\Delta n$ with
energies between $E$ and $E+\Delta E$ is given by a semiclassical expression
related to the volume in phase-space (see (13) below with $s=0$). The factor 
$h$, the size of the elementary cell in this semiclassical approach, is
chosen to contact with quantum theory. Thus (2) suggests to consider for the
number of microstates $\Delta n$ the expression: 
\begin{equation}
\Delta n=\frac{1}{h(1+sE)}\int_{E<H<E+\Delta E}dxdp.
\end{equation}

Note that the integral in (13) coincides with the usual one ($s=0$) because
the functional form of the Hamiltonian $H$ does not change as a function of
the variables $x$ and $p$. The density of states $\rho =\frac{\Delta n}{%
\Delta E}$ becomes

\begin{equation}
\rho =\frac{\rho ^{(o)}}{1+sE},
\end{equation}
$\rho ^{(o)}$ being the usual density function. The partition function for
the Boltzmann distribution is given by

\begin{equation}
Z=\int \frac{e^{-E/T}\rho ^{(o)}}{1+sE}dE,
\end{equation}
where we consider units so that the Boltzman's constant $k=1$. From (15) we
can obtain the thermodynamical quantities. For instance, the internal energy 
$U$ $=T^{2}\partial \ln Z/\partial T$ becomes

\begin{equation}
sU=\frac{Z^{(o)}}{Z}-1,
\end{equation}
where $Z^{(o)}$ stands for the usual function partition ($s=0$). From
(15-16), we have the first correction in the parameter $s$ for the internal
energy :

\begin{equation}
U=U^{(o)}-sT^{2}\frac{\partial U^{(o)}}{\partial T}+O(s^{2}).
\end{equation}
\ 

>From (17), the heat capacity could be computed from the usual definition $%
C_{V}=\partial U/\partial T$. Note that the use of (17) is easy since one
only needs to know the energy of the unperturbed system.

\[
{} 
\]

Recall that equation (17) allows us to compute the first order correction to
the internal energy from the usual expression for the internal energy $%
U^{(o)}$. Moreover, when $U^{(o)}$ increases with temperature, for $s>0$,
the new internal energy is smaller than the usual one.

\[
{} 
\]

For the ideal gas, where $U^{(o)}=\frac{1}{2}NT$ , we have $U\approx \frac{1%
}{2}NT(1-sT)$ and the correction to the heat capacity is $C=\frac{1}{2}%
N\left( 1-2sT\right) $. These results agree with reference [4].

\[
{} 
\]

Now we shall consider $N$ non interacting particles in an external energy
potential like (10). When $s=0$ it is direct to show that

\begin{equation}
Z^{(o)}=T^{\frac{1}{\nu }+\frac{1}{2}}\int e^{-x}x^{\frac{1}{\nu }-\frac{1}{2%
}}dx;\qquad U^{(o)}=\left( \frac{1}{\nu }+\frac{1}{2}\right) NT.
\end{equation}
We note that the density of states can be evaluated directly from (12).
Using (17) and (18) we obtain the internal energy to first order in $s$:

\begin{equation}
U=\left( \frac{1}{\nu }+\frac{1}{2}\right) NT\{1-sT\}.
\end{equation}
While the heat capacity, to this order becomes

\begin{equation}
C=\left( \frac{1}{\nu }+\frac{1}{2}\right) N\left( 1-2sT\right) .
\end{equation}
where the usual harmonic oscillator corresponds to the case $\nu =2$.

\[
\]

For the phonon gas at the regime of low temperature with internal energy $%
U^{(o)}=AT^{4}$, where $A$ is a constant [14], we obtain $U\approx
AT^{4}(1-4sT)$. The corresponding correction to the heat capacity is $%
C=A\left( 4T-20sT^{4}\right) $.

\[
\]

\[
{} 
\]

\begin{section}*{ IV  Conclusions and discussion}
\end{section}

We have studied some systems with noncanonical commutation rules like (1).
This was carried-out using a semiclassical method and the surmise $\hbar
\rightarrow \hbar \{1+sE\}$. With this method we have reproduced the
spectrum of the particle in a quantum well [4] and the harmonic oscillator
[2,3]. Moreover, we have given an explicit equation to determine the
spectrum of a particle in an external energy potential like $\left| x\right|
^{\nu }$ with $\nu $ a positive constant (12).

\[
{} 
\]

Thermodynamical properties of a system with noncanonical commutation rules
can be directly calculated, if we note that our surmise implies a change in
the size of the elementary cell in phase space. We have obtained the general
expression for the density of states (14) and the partition function (15).
to first order in the parameter $s$, we have found an expression for the
internal energy $U$ as a function of the usual one $U^{(o)}$. We have
applied this method to study the ideal gas, the phonon gas and the
noninteracting particle in a potential like $\left| x\right| ^{\nu }$.
Thermodynamical properties were calculated using the Boltzman distribution.
Since we have the density of states (14) in principle we can calculate the
partition function with the Fermi-Dirac or Bose-Einstein distribution.

\[
{} 
\]

As mentioned in the introduction, charge quantization for mesoscopic
circuits can be considered by noncanonical quantization rules like (1)
[7-10]. These physical systems are today the object of research because of
eventual technological applications. Our semiclassical calculation of energy
spectra and thermodynamical properties is useful as a first approach to
these mesoscopic systems. For instance, from the analogy between an electric
circuit and the harmonic oscillator, the spectrum (9) could be related to
the LC quantum circuit with charge discreteness.

\[
\]

As a curiosity, we remark that several experiments in particle physics show
that the fine structure fine $\alpha =e^{2}/c\hbar $ depends on the energy
[15,16]. This is usually interpreted as the change of the electron charge
with the energy parameter. It seems to us that this presents an analogy with
our results, namely, an energy dependence of the constant $\hbar $. This
subject will be studied elsewhere.

Finally, as mentioned in the introduction, the case $\hbar \rightarrow \hbar
e^{sE}$ in (2) is soluble in many cases since here \ $Z=\int e^{-E/T-sE}\rho
^{(o)}dE$ \ and defining the effective temperature $\frac{1}{T^{\ast }}=%
\frac{1}{T}+s$ then we can operate with the formal change $T\rightarrow
T^{\ast }$ in any solvable thermodynamical system.

\[
{} 
\]

Acknowledgments: This work was possible due to the project UTA-Mayor (cc
4723), and FONDAP Matem\'{a}ticas Aplicadas. J.C.F. thanks useful e-mail
correspondence with professor You-Quan Li. We thank the Universidad de la
Frontera (UFRO) because of the facilities furnished to S.M.

\ 

\[
{} 
\]

\newpage

\end{document}